# Cyber-Physical Microservices

An IoT-based Framework for Manufacturing Systems

Kleanthis Thramboulidis, Danai C. Vachtsevanou, Alexandros Solanos
Electrical and Computer Engineering
University of Patras, Patras, Greece

*Abstract*—Recent advances in ICT enable the evolution of the manufacturing industry to meet the new requirements of the society. Cyber-physical systems, Internet-of-Things (IoT), and Cloud computing, play a key role in the fourth industrial revolution known as Industry 4.0. The microservice architecture has evolved as an alternative to SOA and promises to address many of the challenges in software development. In this paper, we adopt the concept of microservice and describe a framework for manufacturing systems that has the cyber-physical microservice as the key construct. The manufacturing plant processes are defined as compositions of primitive cyber-physical microservices adopting either the orchestration or the choreography pattern. IoT technologies are used for system integration and model-driven engineering is utilized to semi-automate the development process for the industrial engineer, who is not familiar with microservices and IoT. Two case studies demonstrate the feasibility of the proposed approach.

## I. Introduction

The fourth Industrial revolution, referred as Industry 4.0, has a tremendous impact on the society and the global economy. IoT technologies along with Cloud computing, data analytics and cyber-physical systems play a key role in this evolution [1][2]. The manufacturing industry is greatly influenced by this revolution since it faces, as authors claim in [3], the challenge of rapidly growing requirement for agile and effective reactivity to the rapidly changing market demands. Moreover, the competitive nature of today's industry forces manufacturing to move towards implementing high-tech methodologies [4]. Thus, it is expected that very soon the IoT-based manufacturing environment will be a reality.

In this context, the main objective of digitalization of production environments is to achieve a new automation paradigm, more flexible, responsive to changes and safe [5]. The service-oriented architectural style has attracted the interest of research and practitioners from the manufacturing domain a long time ago, e.g., [6] and [7], since the identification of the most suitable architecture greatly impacts not only the development process of the system [8] but also its entire life cycle. However, the adoption of research results on SOA is not the expected one in practice. The manufacturing industry is conservative and is expecting for a technology to reach an acceptable level of maturity before its adoption. During that time a new paradigm based on the concept of microservice appeared in SOA and promises to change the way in which software is perceived, conceived and designed [9]. Microservices are the building block of the microservice architecture, that is one of the latest architectural trends in software engineering [10], promising to address several open issues in software development [11].

The potential of this new paradigm has been identified and we claim that the microservice paradigm will have a significant impact on the way future manufacturing systems will be developed. Thus, in this paper we propose the integration of IoT technologies with the microservice architecture and examine alternative scenarios for their exploitation. Based on this, a framework for the exploitation of both technologies, i.e., microservices and IoT, in the manufacturing domain, is briefly presented. However, the investment in traditional technologies, as for example IEC61131 based systems, is huge. There is a need for systems and components that have been developed based on the conventional approach to be integrated and exploited in the new environment. On the other side, the adoption of microservice and IoT technologies in the manufacturing domain will greatly affect the development and operation processes of systems in this domain. Industrial engineers are not familiar with these technologies, which when adopted make the development process too complicated for them. Furthermore, there are also other challenges that manufacturing faces, such as the need to switch from mass production to mass customization, and the strong demand for real-time response at the machine control level. Microservices and IoT technologies that will be adopted in the manufacturing domain should properly address these challenges.

Model-driven engineering (MDE) is used to address the complexity of the development process as well as to get the other benefits of this paradigm in the manufacturing domain. The physical units of the manufacturing plant (referred as plant in this paper) are transformed to intelligent (smart) entities that we call cyber-physical microservices (CPMSs). CPMSs are described using web technologies and are available for discovery and use during the development time of the manufacturing system processes, but also during its operation to have a flexible manufacturing system able to address the challenge of mass customization. Moreover, the modularity at the plant process layer, that is required to address mass customization needs, is increased by modelling manufacturing processes using the microservice architecture. Composite CPMSs are defined by integrating primitive ones. IoT protocols are utilized as glue among the constituent microservices of a plant's composite microservice as well as at the system integration level. Microservices are considered as resources in the manufacturing environment and the Resource Description Framework (RDF) [12] is utilized to have a machine-readable specification for the primitive but also for the composite ones that the plant offers. RDF is also used to capture the domain knowledge in terms of models and meta-models which enrich the framework.

The contribution of this paper is the description of a framework for the exploitation of the emerging microservice architecture in the domain of manufacturing systems. The framework utilizes model-driven engineering to semi automate the utilization of the microservice and IoT related technologies and handle the complexity introduced by these technologies in the engineering of manufacturing systems. The effects of the utilization of these technologies on the architecture of the manufacturing systems, as well as on development process are investigated and discussed. The rest of this paper is structured as follows: Section 2 presents background information and related work. The cyber-physical microservice is defined in Section 3. An implementation approach for the cyber-physical microservice is also presented. Section 4 describes the modeling of the plant process layer using the microservice paradigm and discusses description and discovery of CPMSs as key issues in the modeling of the plant. The paper is concluded in the last section.

## II. BACKGROUND AND RELATED WORK

As authors argue in [13], traditional manufacturing systems are slow in responding to market or supply chain changes and this is mainly due to the fact that they are based on the 5-layer architecture (ISA-95 model). The adoption of recent ICT into the Manufacturing domain will enable real-time response to changes in the factory, the supply chain and customers' needs [13]. To address these requirements, as well as the challenge of integration of the three different disciplines in CPSs [14], we have adopted in this work the architecture shown in Fig. 1. This architecture is based on the adoption of concepts from the cyber-physical and IoT domains [15] and is adapted to the microservices architecture. Our architecture is different from the 5-layer architecture proposed in [4] mainly due to the different understanding of the term cyber-physical. We consider as cyber-physical the lower level of the 5-layer architecture, i.e., the *Smart Connection Level*, that captures the tight integration of the physical with the cyber world. The 5-layer architecture defines the third layer as *Cyber Level*.

Research on Cyber-physical systems plays a key role in the evolution of manufacturing systems. CPSs, which are the essential part for the future internet of things [16], are expected to have a significant impact on the way manufacturing systems are developed and operate. A great number of standards attempt to define a common language and the basic concepts regarding the exploitation of CPS research in manufacturing systems. Authors in [17] present a critical evaluation of international standards with the intention to help academic scholars and industry practitioners to manage challenges that should be addressed in the fourth revolution of manufacturing systems. In [18] authors describe a manufacturing platform that utilizes Internet of Things, cloud computing, big data analytics, cyber-physical systems, and prediction technologies and claim that its application to the production process in the semiconductor domain is promising to achieve the goal of zero defect. Authors adopt RESTful services for horizontal and vertical integration of system components.

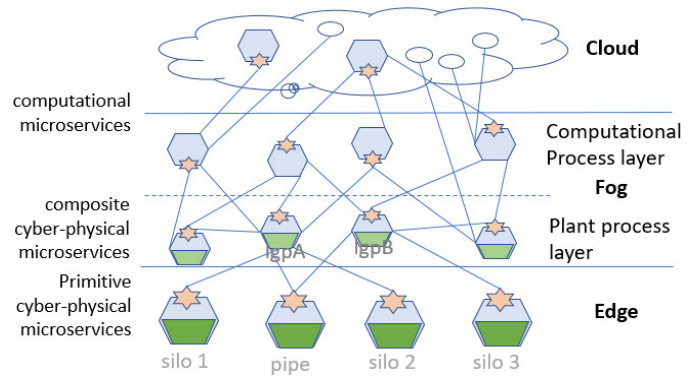

Fig. 1. The Manufacturing system architecture based on Cyber-Physical Microservices.

In [13] authors describe a framework for modeling of cyber-physical manufacturing systems utilizing SOA and web technologies. The framework supports, as authors claim, automatic service discovery, identification and orchestration and utilizes web service integration based on semantics. The framework is adapted to the requirements and constraints of OPC UA (https://opcfoundation.org/about/opc-technologies/opc-ua/). Our work has many similarities with the work in [13] but, a) it introduces the concept of modularity based on cyber-physical microservices, b) uses IoT technologies for the integration of functionalities, and, c) utilizes MDE to automate the development, commissioning and operational phases, to mention the most significant differences.

Modern manufacturing systems should be able to scale and evolve over time to satisfy the changing requirements of the market adopting innovative technologies and designs. The microservice architecture style has emerged and gained a lot of popularity in the industry in recent years. In [19] authors present a prototype platform that supports multiple concurrent applications for smart buildings. The proposed platform utilizes advanced sensor network in a distributed microservices architecture. Authors claim that the use of microservices results to a platform that promises strong scalability, reliability, and ease of evolution regarding hardware resources and finally direct utilization of the immense power of external services available on the Internet. In [20], authors describe an IoT platform for smart city that utilizes the microservices architecture style. A survey of commercial frameworks for the IoT can be found in [21].

The microservice architectural style has been recently adopted by various large companies and is becoming popular since in some cases it appears as the only feasible solution for reducing the growing complexity of systems [8]. A microservice (MS) is defined as a cohesive, independent process interacting via messages [9]. A microservice focuses on a single aspect [33] and can be reused, orchestrated, and aggregated with others [9]. MSs bring simplicity in components management, reduce development and maintenance costs, and support distributed deployments [19]. These characteristics make microservices a promising technology for manufacturing systems. As authors claim in [22], benefits of this technology include among others, increase in agility, developer productivity, resilience, scalability, reliability, maintainability, separation of concerns, and ease of deployment. The results of an empirical study conducted by authors in [8] identify maintenance, scalability, delegation of responsibilities to independent teams, and a better support for DevOps, as main motivations for migrating to microservices. Authors in [20] present their experience regarding benefits of microservices compared to the traditional SOA approaches.

Microservices have already attracted the interest of the research community in the domain of manufacturing systems. Authors in [5] utilize microservices for the construction of a framework to facilitate the integration of simulations into the digital factory. Authors claim that they plan to extend their framework by including other more IoT-friendly protocols like XMPP and MQTT, to facilitate the integration with legacy or heterogeneous solutions. Our work utilizes LwM2M that is implemented over CoAP. The MQTT based implementation of LwM2M allows our approach to also utilize MQTT.

Authors in [23] utilize microservices to propose a collaborative Industry 4.0 platform that enables IoT-based real-time monitoring, optimization and negotiation in manufacturing supply chains. They claim that microservices provide better scalability, better agility and continuous delivery and facilitate new levels of customization, security, workload changes, simplify validation and testing. As service communication they utilize HTTP or the Apache Kafka messaging service. We use in our platform the OMA LwM2M that has several advantages compared to HTTP regarding the manufacturing environment, since our platform considers also the shop floor units and their integration. Instead, in [23] a specific component, the IoT component, is proposed to support communication among IoT devices and the microservices of the proposed platform using MQTT.

An infrastructure for automated deployment of microservices for monitoring purposes is presented in [24]. The author adopts the container-based approach, instead of the hypervisor-based one, and claims that the microservice approach is a promising design paradigm that is tightly bound to the container technology. Authors in [25] utilize microservices to design and implement a dynamic orchestrator in a cloud-based computing environment. The proposed orchestration framework automates the dynamic adjustment of applications to the up and down scalability of cloud resources. This work can be transformed and adapted to our framework to address mass customization needs.

To the best of our knowledge there is no other work that a) adopts and examines the microservice paradigm in the shop floor level of manufacturing systems, and b) considers the CPMS as the key construct of manufacturing systems and IoT as a glue for the integration of CPMSs.

## III. CYBER-PHYSICAL MICROSERVICES

The proposed framework is based on our previous work [26], which utilizes Model Integrated Mechatronics (MIM) in the domain of manufacturing cyber-physical systems. In this transition from mechatronic systems to CPSs we have considered CPSs as extension of mechatronic systems. Arguments on this view can be found in [27]. Two case studies, the liqueur plant [28] and the Gregor chair assembly system [29] are under development and investigation based on the cyber-physical microservice approach.

### A. Microservice-based development of manufacturing CPS

In this work we adopt the microservice paradigm and define the cyber-physical microservice as the key construct of the cyber-physical system. As shown in Fig. 2, the manufacturing cyber-physical system is defined in the cyber-physical microservice layer as a composition of well-defined CPMSs, which are either primitive or composite. A primitive CPMS (PCPMS) is composed by the PCPMS developer who is tightly integrating mechanics, electronics and software with the objective to realize specific functionalities that the PCPMS would offer to its environment. A PCPMS has also, along with its software interfaces, well defined electronic and mechanical interfaces, which differentiates it from traditional microservices of the software domain. All these interfaces are defined using the SysML construct of provided and required interface [26]. AutomationML can also be used for this purpose. The identification and definition of CPMSs for a domain, e.g., assembly systems, and a specific application is a great challenge. It requires very good knowledge of the domain, so as to properly capture a) the domain knowledge into domain MSs, and b) the application knowledge into application MSs.

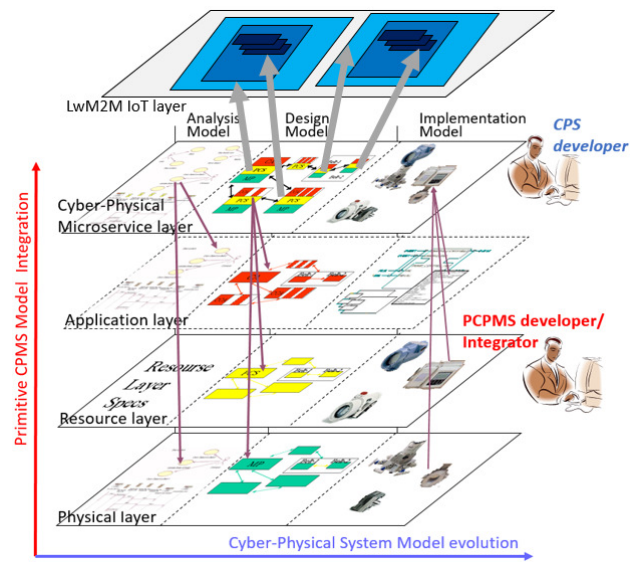

Fig. 2. A microservice and IoT-compliant architecture for Cyber-Physical manufacturing systems.

### B. Primitive and Composite cyber-physical microservice

A CPMS implements its functionality by a close integration of physical and cyber artefacts. It offers a specific and narrowly defined physical functionality, such as heat and mix, enriched by cyber artifacts and is deployed on the plant platform as an independent service but within the context defined by the constraints imposed by its mechanical part. We discriminate microservices into primitive and composite microservices, as shown in fig. 3, which captures part of the UML4IoT profile [15] that has been extended to support the CPMS paradigm. CPMSs interact with their environment through well-defined ports (*CPMSPort*) that are characterized by their provided (*itsProvidedIf*) and required (*itsRequiredIf*) interfaces.

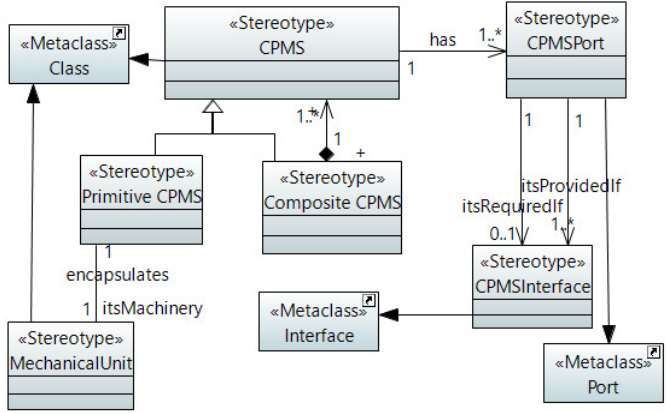

Fig. 3. Part of the UML4IoT profile that catpures Cyber-physical Microservices.

*Primitive cyber-physical microservice* is the microservice that encapsulates a physical artefact by adding intelligence on it and transforming it to a smart entity that is able to either process, or transport, or store energy, material, parts, sub-products and products. Typical examples of PCPMSs are the smart silo and smart pipe of the Liqueur Plant case study and the three Robot arms and the two workbenches of the Gregor chair case study. PCPMSs operate on physical objects, e.g., assembly parts, and transform their state. A PCPMS (*PrimitiveCPMS*) has its own dedicated mechanical part (*MechanicalUnit*), that is completely under its control. However, the developer may define specific parameters of the physical part to be readable from the environment. This is supported by the provided interface of the CPMS (*itsProvidedIf*). Required interfaces (*itsRequiredIf*) capture the dependencies among MSs. Microservice Interfaces can be implemented by the Interface construct of OO IEC 61131 [30], which is supported by commercial tools, e.g., CoDeSys 3. PCPMSs are implemented as real world entities which we call Industrial Automation Things (IATs). An IAT, which is an edge node, may implement more than one CPMS, as is the case of the processing unit IAT of the Festo MPS system [31], which hosts three CPMSs, the rotating disk, the drilling machine and the checking machine CPMSs.

*Composite cyber-physical microservice* is the microservice that offers its functionality by utilizing (directly or indirectly) the functionality of at least one PCPMS. Composite CPMSs implement functionality for either processing, and/or transporting, and/or storing material, parts, sub-products and products. The operation of composite cyber-physical MSs depends on the availability of the utilized primitive cyber-physical MSs with which it is coupled in time, that is another differentiation from software microservices. Typical example of CCPMSs are plant processes as for example the liqueur generation processes of the Liqueur Plant case study.

The effective exploitation of the microservice concept and containers at the field device level, where real-time constraints should be met, is a great challenge for the researchers of the industrial automation domain, since existing microservice containers, such as Kubernetes (https://kubernetes.io/) or Docker [32], introduce high latency in integration. To this direction, frameworks for the embedded domain and automotive systems are soon expected [33]. Support from these frameworks to low latency communication mechanisms, such as the operating system's IPC transports, should provide a solution to this problem in case of MS integration in the same node. These frameworks will allow the modeling of smart objects such as silo and robot arms as CCPMSs by utilizing the concept of microservice at the field device level. In this case drivers of the various parts of a mechanical unit, e.g., valve, heater, mixer, of the silo unit, will be modeled as PCPMSs resulting to a more flexible smart object implementation. The orchestration inter-service interaction pattern is utilized for the construction of the controller as a composite cyber-physical microservice.

*C. Microservice orchestration overhead*

Microservices interact via messages. In our prototype implementation the LwM2M IoT protocol is adopted for the interaction of microservices. Both the choreography and the orchestration pattern can be used for the integration of microservices. A classical example of the use of the orchestration pattern that resembles to the centralized control approach is the modeling of a plant process such as the liqueur generation process. A classical example of the choreography patterns, that resembles to the distributed control approach, is the implementation of the assembly process in the Gregor chair case study.

As an indication of the performance overhead that is introduced by the different integration mechanisms for microservices, Table I presents performance measurements expressed in ms, regarding the latency introduced by communication mechanisms adopted in various levels of microservice integration. An interesting study of the performance of HTTP compared with the one of Modbus is given in [34]. Authors argue that HTTP can be considered as a worth evaluating alternative for soft real-time networked control systems. It should be noted that CoAP, which is used in this framework, is based on UDP and thus eliminates the overhead introduced by HTTP and Modbus TCP which are based on TCP. Table I captures the round-trip time for the Execute operation of the LwM2M, with all its data to be based on 1000 measurements for each configuration scenario. The basic two node configuration scenario (2N) is composed of: a) the plant process composite CPMS (liqueur generation process type A - lgpA) that was deployed on a fog node, i.e., an Intel i5-4590 CPU running at 3.3 GHz and 8 GB of DDR3 RAM, running Ubuntu 16.10 64-bit OS and Node.js v6.11.1, and b) the smart silo IAT controller implemented as primitive CPMS using Node.js as runtime platform and deployed on an edge node, i.e., a Raspberry Pi 3 Model B connected to a router. Node.js is single threaded so it perfectly matches with the scan cycle model used in manufacturing. Node.RED (https://nodered.org/) was used for a user-friendly specification of the composite CPMS, following a process-driven approach [15].

TABLE I. LATENCY IN CYBER-PHYSICAL MICROSERVICE INTEGRATION

| | IPC | | Network Communication | | | | | | | |
|---|---|---|---|---|---|---|---|---|---|---|
| | Unix socket | Pipes | LwM2m | | | amqp | | UDP Sockets | | |
| | | | 1N | 2N | 2N+DC | 2N | 2N+DC | 1N | 2N | 2N+DC |
| avg | 0.21 | 0.28 | 8.74 | 4.9 | 5.05 | 2.17 | 2.2 | 0.54 | 0.5 | 0.69 |
| min | 0.10 | 0.08 | 3.52 | 2.01 | 2.1 | 0.97 | 1.11 | 0.2 | 0.28 | 0.37 |
| max | 23.23 | 9.34 | 138.6 | 91.16 | 89.19 | 71.5 | 67.8 | 38.86 | 21.01 | 32.66 |
| stdev | 0.82 | 0.78 | 7.72 | 5.4 | 4.99 | 2.6 | 2.73 | 1.34 | 1 | 1.41 |

Unix domain sockets and pipes were utilized for the integration of microservices on the same node, as is the case of implementing the smart silo IAT controller as a CCPMS, while LwM2M, advanced message queuing protocol (amqp) (https://www. amqp.org/) and UDP sockets are used as network communication mechanisms. The table also captures the overhead introduced by the network communication mechanism in the case that both client (smartSilo) and server (lgpA) microservices are deployed on the edge node (1N). In this case, the RPi, introduces significant performance overhead regarding the execution of the protocol stack of the process side. That is why the round trip for the one node (1N) scenario is higher than the two nodes (2N) one. Moreover, the difference is higher for the lwM2M and amqp compared to UDP as it is expected. The corresponding measurements for lwM2M for the 1N scenario using as node the fog node present an avg. 1.77 ms with min. 0.52 and max. 17.7 ms.

LwM2M appears to introduce a higher latency compared to amqp, even though it is based on UDP while the latter is based on TCP. Our estimation is that the node.js support for LwM2M is not optimized since the corresponding java LwM2M implementation for the client MS presents much better latency in its integration with a node.js server MS. (avg: 3.34, min: 1.88, max: 49.14, stdev:2.81). To have a measure of comparison for the overhead that the various MS integration mechanisms introduce, we note that the function call overhead is in the range of a few nanoseconds (ns). This is a convincing argument for avoiding the use of microservices at very low level of functionality, such as function blocks or even operations, such as heat and mix in the case of smart silo. Table I captures also, in columns indicated as 2N+DC, the overhead introduced by the use of the Docker container. These measurements are for a 2N scenario where Docker 17.05.0-ce (build 89658be) is used as container at the fog node. Our proposal is to avoid the use of microservices and microservice containers, such as Kubernetes or Docker, for the development of the software part of the primitive CPMS since this introduces a high latency in integration. Instead cyber-physical components of the industrial system, i.e., primitive CPMSs, such as smartSilo, can be implemented using traditional well proven technologies and expose their functionality as microservices using IoT technologies, as presented in [35].

### D. The CPMS Architecture

For a traditional plant machinery of the manufacturing domain to be integrated into the microservice and IoT-based manufacturing environment, it should be transformed into a CPMS that provides a RESTful interface. We have adopted the OMA LwM2M application layer protocol, which is implemented on top of CoAP, (an MQTT based implementation also exists) to provide an IoT-compliant interface for the CPMS, as shown in figure 4. A crucial aspect for the success of the IoT is the interoperability challenge [36]. Microservices will be developed by different vendors with different data models and interfaces and this heterogeneity should be resolved for succesfull IoT realizations [16]. Authors in [3] present a layer to obtain interoperability between the physical and cyber layers of industrial CPSs. They position the controller in the physical layer of the CPS, considering in this way our cyber-physical microservice as belonging to the physical layer of the CPS. We adopt IPSO objects to address interoperability requirements among different microservices. However, since it is not expected to have a convergence on a single IoT communication protocol as claimed in [36], protocol translators will provide a solution to address interoperability requirements with non-IPSO compliant systems.

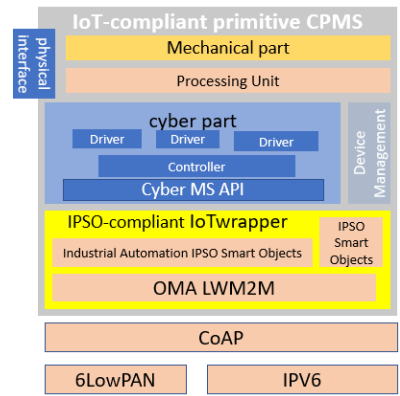

Fig. 4. Architecture of the IoT-compliant cyber-physical microservice.

The IoTwrapper is the software layer that transforms the legacy interface of the cyber part of the plant machinery to an IoT-compliant one. It transforms the conventional plant machinery to an IoT-compliant microservice. We found the adaptation process too complicated for the Industrial Engineer and this was the motivation to use MDE to automate its construction. For the specification of the IoT-compliant interface of the CP microservice, the LwM2M provides an object model that is based on the concept of Resource. This model focusses only on the modelling of the interface. On the other side, the traditional cyber-physical component, whose part is the plant machinery, has been specified with an object model that also specifies its interfaces. UML and SysML, the de-facto standards for software and system engineering, are commonly used for such specification. Thus, we have two models; one focuses only on the IoT-compliant interface description, and the other on the modeling of the whole cyber-physical MS including its interface, which cannot however be specified in an IoT-compliant way.

To address the above problem, we have defined the IoT layer on top of the cyber-physical MS layer of the extended MIM Architecture, as shown in Fig. 2. For the definition of the modeling space of this layer the basic constructs of the LwM2M object model were formalized using UML as shown in Fig. 5, that captures the LwM2M communication protocol interface. In this way projecting the cyber-physical MS layer model of cyber-physical manufacturing system to the IoT layer (Fig.2) we get the IoT compliant interface for the cyber-physical microservices of the system, as well as, for the system as a whole. UML was adopted as base for the transformation process between the two layers, and the UML4IoT profile was extended to implement this projection.

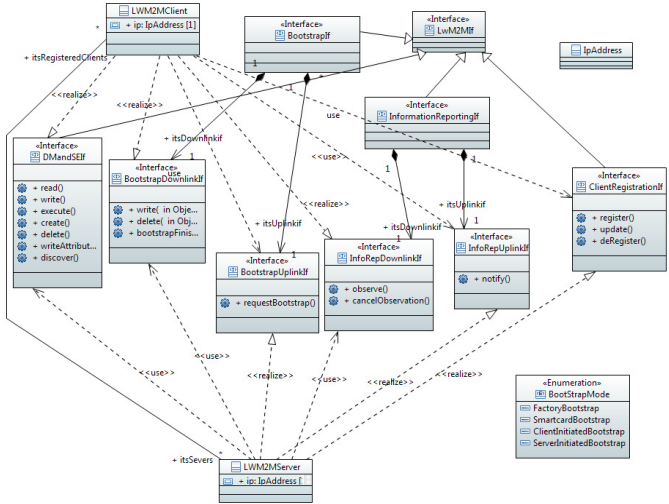

Fig. 5. Model of the LwM2M communication protocol interface adopted at the IoT layer of architecture of the cyber-physical manufacturing system.

## IV. THE PLANT PROCESS AS MICROSERVICE

Plant processes can be defined adopting the choreography or the orchestration pattern. We have applied the choreography pattern for the Gregor chair assembly system case study to define the assembly processes and the orchestration pattern for the Liqueur Plant case study. Plant processes, such as the liqueur generation processes of various types, e.g., lgpA and lgpB of the Liqueur Plant case study [28], utilize directly or indirectly functionality provided by primitive CPMSs as well as computational MSs, to provide a higher layer functionality required at the process level of the plant, as shown in Fig. 1. Thus, plant processes are realized as composite CPMCs. Chunks of functionality at the plant process layer that involve more than one CPMS are also modeled as CPMSs to have a modular and flexible process layer implementation. The CPMS *transfer liquid from source silo to destination silo* is a classic example of a composite CPMS.

A plant process is specified as an orchestration of functionalities offered by plant units, thus, the corresponding CCPMS that realizes the plant process is defined as orchestration of CPMSs. Several notations are used for service orchestration with the goal to be usually twofold, flexibility and responsiveness. Flexibility, which is a highly desirable feature in smart factories [37], means to adapt to changing requirements in production, and responsiveness to respond to the physical plant stimulus meeting the deadlines that the plant units impose. The objective of the proposed approach is to fulfill both requirements. Responsiveness is addressed at the primitive CPMS level by encapsulating the mechanical unit control and coordination logic in the MS level, i.e., in the PCPMS, close to the physical plant unit. Flexibility is achieved by several means. As a first step, plant processes are implemented as dynamically deployable MSs, which are executed in a MS container that supports run-time reconfiguration, e.g., OSGi [15]. Moreover, plant processes may be defined without any reference to specific services provided by the plant. This allows a plant process, i.e., a composite CPMS, to dynamically acquire at deployment and even at run-time, the available CPMSs, which are required to fulfill its goals, i.e., its requested service specs.

Based on this scenario, the plant process developer defines the plant independent model (PIM) for the process, i.e., they specify the process in a plant independent manner. PIM specifies the operations that should be performed without using specific operations that depend on the plant configuration. For example, operations such as fill, empty and transfer, for the case of liqueur plant, have to do with the plant configuration and are not included in the PIM model. These operations will be inserted in the model in the next phase when the PIM will be transformed to a plant specific model (PSM), i.e., during the time a requested service spec of the PIM is resolved to a specific provided service, an action that customizes the PIM to the specific configuration of the plant. If, for example, the requested heating service spec is not supported on the current silo, its content should be transferred to the silo that provides this service, an action that transforms the PIM to PSM.

### A. PIM to PSM transformation

The transformation of the PIM to PSM can be performed manually by the control engineer or may be automated by the framework. The framework supports this operation through a service discovery mechanism. This mechanism can be utilized either for a static assignment of provided services or a more flexible dynamic one. In the case of dynamic assignment of services, the system will check for the availability of primitive CPMSs providing the physical operations and satisfying the requested service specs and the prerequisites of using them. Then, it will instantiate the process MS reserving the required CPMSs. An alternative is for the system to postpone the reservation of resources up to the time they are required. This functionality of the framework supports a better use of the plant's resources and allows a more flexible process implementation. The MS description is a prerequisite for the realization of the PIM to PSM transformation.

### B. Cyber-physical MS description and discovery

A primitive cyber-physical MS, such as the SmartSilo, has several exposed resources provided as services, as for example heat and mix. These services will be utilized for the realization of composite cyber-physical MSs, as for example the *transfer liquid from source silo to destination silo*. Moreover, for the framework to support service discovery during the development time but also during run-time an efficient description is required for the MS and the operations that the MS implements. We have complemented the IPSO smart objects by RESTdesc descriptions of the offered plant operations and the MSs' states. RESTdesc is a machine-interpretable functional service description format for REST APIs [38] that exploits HTTP vocabulary and Notation3 (https://www.w3.org/TeamSubmission/n3/) to enable the machine to discover and consume Web services based on links [39]. N3 extends the Resource Description Framework (RDF). It is based on Statements, which are triples consisting of a Resource, a Property and the value of the Property, represented by URIs and serving as subject, predicate and object, respectively. For example, the triple *local:heat a lps:Service* of Fig. 6 defines heat as a Service (*a* is an abbreviation of N3 for the *rdf:type* property) and the *rdfs:label* instance of Property is used to define a human readable name for the resource. Properties are also used to express attributes of a resource or a relationship between two resources. RESTdesc descriptions include a set of preconditions and a set of postconditions, indicating that if the preconditions in the antecedent are true for a specific substitution of the variables, then an HTTP request will be feasible for the realization of a service by using URIs or request bodies associated with the same substitution. MSs' states are described by a mechanism that allows RESTdesc to capture states, which was introduced in [40] and extended in [41]. N3 statements may provide information about the functionality of a service, its inputs and outputs and information about Quality of Service (QoS) characteristics. For example, all heating services should have a common label "Heat", defined by a corresponding ontology, but possibly different levels QoS regarding the maximum allowed heating temperature or the types of material that can be processed. Fig. 6 captures part of the description for a heating service which is labelled accordingly and has defined QoS characteristics, i.e. it accepts only input temperatures expressed in Celsius unit, it can heat up to 70º Celsius and it is destined for materials of liquid type.

The framework supports the discovery of MS using a service repository where the CPMSs of the manufacturing plant are automatically registered by their hosting devices (IATs). The CoRE resource directory [42] defined by the IETF CoRE Working Group is adopted in this work. It enables methods for discovering a resource directory (RD), as well as registering and looking up resource descriptions. Although in the manufacturing domain sleeping nodes and intermittent connection to constrained network are not the case, direct discovery of resources provided by devices may not be feasible in most smart environments [43]. The CoRE RD targets resource-constrained devices used in M2M applications and surpasses the problems that direct discovery imposes, by employing an RD which hosts accessible descriptions of resources held on servers [42]. We use the Cf-RD resource directory implementation of the californium.tools repository (https://github.com/eclipse/californium.tools) to be aware of the devices and services of the manufacturing plant.

```
1  PREFIX local: <http://ss4.ece.upatras.gr/>
2  PREFIX lps: <http://ssegvml.ece.upatras.gr/LiqueurPlantSystem#>
5  PREFIX xsd: <http://www.w3.org/2001/XMLSchema#>
6
7  local:material rdf:type lps:AllowedMaterial;
8                 lps:hasMaterialType dbpedia:Liquid.
9  local:unit rdf:type lps:AllowedUnit;
10             lps:hasUnitType dbpedia:Celsius.
11 local:maxTemp rdf:type lps:MaxTemperature;
12               lps:hasValue "70"^^xsd:double;
13               lps:hasUnit local:unit.
14 local:heat a lps:Service;
15             rdfs:label "Heat"@en;
16             lps:QoS local:unit,local:maxTemp,local:material.
```

Fig. 6. RESTdesc of heat service of the smartSilo cyber-physical MS (part of).

Each device hosting CPMSs accesses the RD and sends a POST request through the registration interface. The message payload contains the list of resources offered by the device in the CoRE Link Format as well as the semantic and dynamic state descriptions of the provided resources. The RD lookup and update mechanisms allow the search and discovery of the exposed resources and the access to up-to-date information concerning resource descriptions. In the Liqueur Plant case study, its components, such as the smart silo, and smart pipe, register to the RD once activated and publish lists of the plant operations they provide, e.g. fill, heat, mix, along with their RESTdesc descriptions. The development environment or an agent, for the case of operation-time discovery, accesses the descriptions and looks for resources that offer the desired functionality for the realization of a composite CPMS, such as the liqueur of type A generation process (lgpA). The SPARQL query language for RDF (https://www.w3.org/TR/rdf-sparql-query/) enables the filtering of services which meet the process requirements. For example, during the development process of lgpB, the control engineer performs queries to identify Heat services with specific QoS characteristics, to specify and potentially utilize the entities that provide these services. Fig. 7 shows a SPARQL query for discovering heating services for liquid, with maximum allowed heating temperature greater than 50º Celsius.

```
1  PREFIX rdfs: <http://www.w3.org/2000/01/rdf-schema#>
4
5  SELECT ?service
6  WHERE   { ?service a lps:Service;
7                     rdfs:label 'Heat'@en;
8                     lps:QoS/lps:hasMaterialType dbpedia:Liquid;
9                     lps:QoS ?maxTemp.
10           ?maxTemp a lps:MaxTemperature;
11                    lps:hasUnit/lps:hasUnitType dbpedia:Celsius;
12                    lps:hasValue ?value
13                    FILTER(?value>=50).
14         }
```

Fig. 7. Example query for the discovery of heat CPMS with specific QoS.

## V. CONCLUSION

The potential of exploiting the microservice architecture along with IoT in the cyber-physical manufacturing systems domain is examined. A framework that exploits these technologies and utilizes MDE to simplify the development process is described. The framework has the cyber-physical microservice as a key construct for the modeling of the system. Performance measurements show that the application of the microservice architecture based on software microservices technologies, i.e., microservice containers and traditional microservice integration protocols introduce a high latency in the level of the cyber-physical component, i.e., smart machinery level, that is usually not acceptable in the industry. Traditional technologies can be used for the implementation of the smart machinery in the form of CPMS and expose its functionality through IoT. On the other side, microservices offer great flexibility at the plant process layer and are considered as a promising technology for manufacturing systems in the context of Industry 4.0. Thus, and taking into account that containers are soon expected for the embedded domain, the CPMS is considered as the key construct for the modular development of flexible manufacturing systems.

ACKNOWLEDGEMENTS

The Authors would like to thank the anonymous reviewers for their comments that resulted in an improved version of the paper.